\documentclass[twocolumn]{aa}
\usepackage{graphicx}
\usepackage{txfonts}
\usepackage{natbib}
\usepackage{slashbox}
\def\aa{A\&A }                  % A&A
\def\apj{ApJ }                  % ApJ
\def\apjs{ApJS }                % ApJS
\def\aj{AJ }                    % AJ
\def\mnras{MNRAS }              % MNRAS
\def\aap{AAP }                  % AAP
\begin{document}
   \title{Observational constraints on models for the
interstellar magnetic field in the Galactic disk}

   \author{H. Men
		  \inst{1, 2}
		  , K. Ferri\`ere\inst{2}
	  \and
	  J. L. Han\inst{1}
		  }

   \offprints{K. Ferri\`ere, ferriere@ast.obs-mip.fr}

   \institute{National Astronomical Observatories, Chinese Academy of Sciences,
			  Jia 20 DaTun Road, Chaoyang District, Beijing, 100012, China\\
   \and Observatoire Midi-Pyr\'en\'ees, Universit\'e Paul Sabatier Toulouse 3,
		CNRS, 14 av. Ed. Belin, 31400 Toulouse, France\\
			 }

   \date{Received  / Accepted }

\titlerunning{Observational constraints on models for the Galactic magnetic field}
\authorrunning{H. MEN et al.}

\abstract
{}
{Our purpose is to place firm observational constraints on the three
  most widely used theoretical models for the spatial configuration
  of the large-scale interstellar magnetic field in the Galactic disk,
  namely, the ring, the axisymmetric and the bisymmetric field models.
}
{We use the rotation measures (RMs) of low-latitude Galactic
  pulsars and combine them with their dispersion measures and
  estimated distances to map out the line-of-sight component of the
  interstellar magnetic field in the near half of the Galactic disk.
  We then fit our map of the line-of-sight field to the three aforementioned
  theoretical field models and discuss the acceptability of each fit,
  in order to determine whether the considered field model is allowed
  by the pulsar data or not.
}
{Strictly speaking, we find that all three field models are ruled out
by the pulsar data. Furthermore, none of them appears to perform
significantly better than the others.
From this we conclude that the large-scale interstellar magnetic field 
in the Galactic disk has a more complex pattern than just circular, 
axisymmetric or bisymmetric.
}
{}


\maketitle

\section{Introduction}

  The interstellar magnetic field of our Galaxy has been the object of
  intense investigation since the early 1980s. Different observational
  methods (e.g., based on synchrotron emission, Faraday rotation,
  Zeeman splitting, polarization of starlight, polarization of
  dust infrared emission) provide information on the magnetic field in
  different interstellar regions. Faraday rotation of Galactic pulsars
  and extragalactic linearly polarized radio sources make it possible
  to directly trace the magnetic field in ionized regions. In
  practice, one measures the so-called rotation measure (RM), defined
  by
   \begin{equation}
	{\rm RM} = 0.81  \int _{0}^{d} n_{\rm e}\ B_{||}\ {\rm d}s \ \ \ {\rm rad \ m^{-2} }\ ,
   \end{equation}
   where $n_{\rm e}$ is the free-electron density (in ${\rm cm
   ^{-3}}$), $B_{||}$ is the line-of-sight component of the magnetic
   field (in $\mu{\rm G}$) and $d$ is the distance to the radio
   source (in ${\rm pc}$).  Pulsars present a number of advantages
   when used as probes of the interstellar magnetic field. In
   particular, they are highly linearly polarized, they have no
   intrinsic rotation measure and their distances can be estimated
   reasonably well. Moreover, the RM of a pulsar divided by its
   dispersion measure (DM),
   \begin{equation}
	 {\rm DM} = \int _{0}^{d} n_{\rm e}\ {\rm d}s  \ \ \ {\rm cm ^{-3} \ pc}\ ,
   \end{equation}
  directly yields the $n_{\rm e}$-weighted average value of $B_{||}$
  along its line of sight,
   \begin{equation}
	 \overline{B_{||}} = 1.232\ \frac{\rm RM}{\rm DM} \ \ \ \mu{\rm G}\ .
	 \label{Eq_B_par_calc}
   \end{equation}

   We now know that the interstellar medium (ISM) is highly inhomogeneous
   and that the interstellar magnetic field has an important turbulent
   component. For this reason, neighboring pulsars may have significantly
   different values of RM and DM, and a plot RM {\it versus} DM will
   generally exhibit a large scatter.
   However, if one considers a Galactic region larger than the scale of
   the turbulent field and containing enough pulsars for statistical
   purposes, one can infer the large-scale (or regular) component of
   $\overline{B_{\parallel}}$ in that region from the slope of the mean
   DM-RM relation \citep{RL94}:
   \begin{equation}
	 \left< \overline{B_{||}} \right> = 1.232 \ \left <
			 \frac{\rm d\ RM}{\rm d\ DM } \right> \ \ \ \mu{\rm G} \ .
	 \label{Eq_B_par_obs}
   \end{equation}

  Various theoretical models have been proposed to describe the spatial
  structure of the large-scale magnetic field in the Galaxy.
  First and foremost are the ring model, the axisymmetric or axisymmetric
  spiral (ASS) model, and the bisymmetric or bisymmetric spiral (BSS) model.
  According to the galactic dynamo theory, ASS fields would be easiest
  to amplify under typical galactic conditions \citep[e.g.,][]{RSS85,
  Ferriere00}, whereas BSS fields could possibly be excited in the presence
  of an external disturbance, such as a companion galaxy \citep{moss95,moss96}.
  On the other hand, the primordial field theory naturally leads to BSS
  fields \citep{Howard97}.

  In principle, RM studies are ideally suited to establish the overall
  structure of the Galactic magnetic field.
  However, the different RM studies performed so far yield contradictory
  results: some favor a ring field \citep{RK89, RL94, vallee05},
  others an axisymmetric or ASS field \citep{vallee91, vallee96},
  and others a bisymmetric or BSS field \citep{SK80, HanQiao94, ID99, Han06}.
%%The work which carefully studied and compared the three models with the
%%pulsar data had been done almost ten years ago \citep{HanQiao94, ID99}.
  Moreover, although all these studies conclude with a preferred field model,
  none of them has seriously considered the possibility that more than
  one model is allowed by the RM data or, alternatively, that none of
  the three basic models alone can account for the data.
  Hence the question we would like to address in this paper:
  which among the ring, axisymmetric and bisymmetric models can clearly 
be accepted on the grounds that it is consistent with the RM data, 
and which model should clearly be rejected on the grounds that it fails
to provide a good fit to the data.

  In recent years, numerous pulsars were discovered in the near half
  of the Galactic disk and many of them had their RM measured.
  At the present time, among the $\sim 1800$ known pulsars,
  690 have measured RMs and, among the latter, 524 are located at low
  Galactic latitudes ($|b|<10^\circ$). Pulsars with measured RMs now
  provide a reasonably good coverage of the near half of the Galactic
  disk. Furthermore, pulsar distances can now be estimated with fairly
  good accuracy thanks to the improved free-electron density model of
  \citet{NE2001} (known as the NE2001 model). The new measurements
  enable one to investigate the configuration of the Galactic magnetic
  field over a much larger region and with much more confidence than
  previously feasible.

  In Sect.~2, we present the three basic theoretical models for the
  interstellar magnetic field in the Galactic disk.
  In Sect.~3, we describe the procedure used to bin the pulsar data
  and to map out the distribution of $B_{||}$.
  In Sect.~4, we fit our map of $B_{||}$ to each of the three field models,
  and we discuss how good the fits are at reproducing the pulsar data.
  In Sect.~5, we summarize our results and conclude our study.

\section{Description of the field models}
   \label{description and requirment}

Throughout this paper, the Galactocentric cylindrical coordinates 
are denoted by $(r, \theta, z)$, and the distance from the Galactic 
center (GC) to the Sun is set to $r_\odot = 8.5 \ {\rm kpc}$.

\begin{figure*}[!htb]
	  \centering
     \includegraphics[angle=-90,width=6cm]{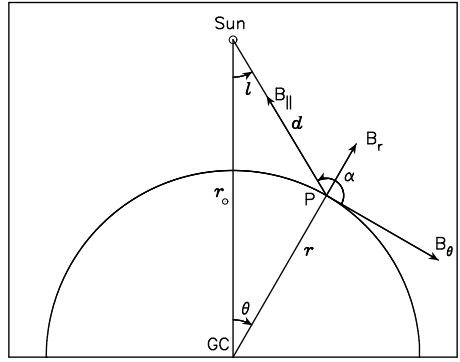}
	   \caption{\label{Fig.1}
		Schematics showing the geometrical variables associated with
		a pulsar P (see main text for the exact definitions).}
	\end{figure*}

In general, the horizontal position of a given pulsar P can be defined either
by its distance from the GC, $r$, and its Galactic azimuthal angle 
$\theta$ (which increases clockwise from $\theta=0$ along the line 
segment GC-Sun), or by its distance from the Sun, $d$, and its Galactic 
longitude, $l$ (which increases counterclockwise from $l=0$ along 
the line segment Sun-GC).  Another useful angular coordinate is the angle 
$\alpha$ between the azimuthal direction at P and the vector P-Sun, 
such that $\alpha=\theta+l+\frac{\pi}{2} $ (see Fig.~\ref{Fig.1}).  

% _______________________ Fig 1 ____________________________

%____________________________________________________________

Because Galactic differential rotation efficiently stretches magnetic field
lines in the azimuthal direction, $\left<B_{\theta}\right>$ dominates over 
both $\left<B_r \right>$ and $\left<B_z \right>$.
Moreover, all the pulsars selected for the present work lie at
$\left| b \right| < 10 ^{\circ}$ and reside in the Galactic disk. 
There, the large-scale magnetic field is nearly horizontal
\citep[e.g.,][]{RSS85, Beck96}, so that $|\left<B_z \right>| \ll
|\left<B_r \right>|, \ |\left<B_\theta \right>|$ . 
In addition, projecting an already small $\left<B_z \right>$ onto 
the line of sight to a pulsar further reduces its contribution 
by a factor $|\sin b| \ll 1$.
Under these conditions, the line-of-sight component of the large-scale 
magnetic field depends only on its radial and azimuthal components,
and is related to them through
        \begin{equation}
          \left< B_{||} \right> = \left<B_r\right> \, \sin \alpha
                + \left<B_{\theta}\right> \, \cos \alpha \ .
          \label{Eq_B_para}
        \end{equation}

We now present the three theoretical field models.

\subsection{Ring model}

In the ring model, the large-scale magnetic field points everywhere in
the azimuthal direction, so that its radial component vanishes:
   \begin{equation}
     \left<B_r\right>=0 \ .
   \label{Eq_ring_Br}
   \end{equation}
Its azimuthal component is constant along circles, i.e., independent 
of $\theta$, but it can vary with $r$\,:
   \begin{equation}
	 \left<B_\theta\right> = \left<B_\theta\right>(r) \ ,
   \label{Eq_ring_Btheta}
   \end{equation}
and it can even change sign along a Galactic radius. 
As a matter of fact, all RM studies leading to a ring model have found 
reversals in $\left<B_\theta\right>$ \citep{RK89, RL94, vallee05}.
It should be noted that the ring model constitutes a particular case
of the axisymmetric model.

\subsection{Axisymmetric model}

In the axisymmetric model, $\left<B_r\right>$ and $\left<B_\theta\right>$ 
are both independent of $\theta$ and vary only with $r$\,:
   \begin{equation}
	 \left<B_r\right> = \left<B_r\right>(r) \ ,
   \label{Eq_AS_Br}
   \end{equation}
   \begin{equation}
	 \left<B_\theta\right> = \left<B_\theta\right>(r) \ .
   \label{Eq_AS_Btheta}
   \end{equation}
Here, too, $\left<B_\theta\right>$ can reverse sign with $r$.
Such sign reversals were found in RM studies favoring an ASS
magnetic field \citep{vallee91, vallee96}.
Interestingly, reversals in $\left<B_\theta\right>$ are also consistent 
with dynamo theory, which can produce them under certain conditions, 
e.g., when the magneto-ionic disk has a particular shape and thickness 
and the seed field itself has strong reversals \citep{PSS93} 
or when the Galactic rotation rate decreases not only with radius 
but also with height \citep{Ferriere00}.

\subsection{Bisymmetric model}

In the bisymmetric model, $\left< B_r \right>$ and $\left<B_\theta\right>$
have a simple sinusoidal dependence on $\theta$, which can be written
in the form
	\citep{Berkhuijsen97}:
   \begin{equation}
	  \left<B_r\right>=b_r(r) \ \sin(\theta-\phi(r)) \ ,
	  \label{Eq_BSS_Br}
	\end{equation}
	\begin{equation}
	  \left<B_{\theta}\right>=b_{\theta}(r) \ \sin(\theta-\phi(r)) \ ,
	  \label{Eq_BSS_Btheta}
	\end{equation}
	where $b_r(r)$ and $b_{\theta}(r)$ are the maximum amplitudes of
	$\left<B_r\right>$ and $\left<B_{\theta}\right>$, respectively, 
        and $\phi(r)$ is the azimuthal phase.
	Both $\left<B_r\right>$ and $\left<B_\theta\right>$
        can reverse sign with $r$.       
	The magnetic pitch angle is defined as
	\begin{equation}
	  \tan p(r)=\frac{\left<B_r\right>}{\left<B_{\theta}\right>}
			   =\frac{b_r(r)}{b_{\theta}(r)} \ ;
	  \label{Eq_BSS_p}
	\end{equation}
	it is positive (negative) if the magnetic field spirals out clockwise
	(counterclockwise) or spirals in counterclockwise (clockwise).

\section {Mapping of $\left< \overline{B_{||}} \right>$}

   To date, there are 690 pulsars with measured RMs \citep{Hamilton87,
   RL94, Qiao95, van97, HMQ99, Crawford01, Mitra03, Weisberg04, Han06, NJKK08}.
   Among these pulsars, we selected those that lie at low Galactic
   latitudes ($\left| b \right| < 10 ^{\circ}$) and have reliable RMs
   (error on ${\rm RM} < 25\ {\rm rad\ m^{-2}}$).
   This left us with 482 pulsars.
   For the distances and DMs of our selected pulsars, we used the values
   given in the ATNF Pulsar Catalog (\citealt{Manchester05},
   see http://www.atnf.csiro.au/research/pulsar/psrcat).
   Pulsar distances in this catalog were estimated with the help
   of Cordes \& Lazio's \citeyearpar{NE2001} NE2001 model for the
   free-electron density; for pulsars located in the inner Galaxy,
   individual distances are typically uncertain by $\sim20\%$, but
   the relative distances of neighboring pulsars have a much lower
   uncertainty.  Pulsar DMs, for their part, are known with good accuracy
   (error on ${\rm DM}$ generally $< 1\ {\rm cm ^{-3} \ pc}$).

\begin{figure*}[!htb]
\centering
\includegraphics[angle=-90,width=12cm]{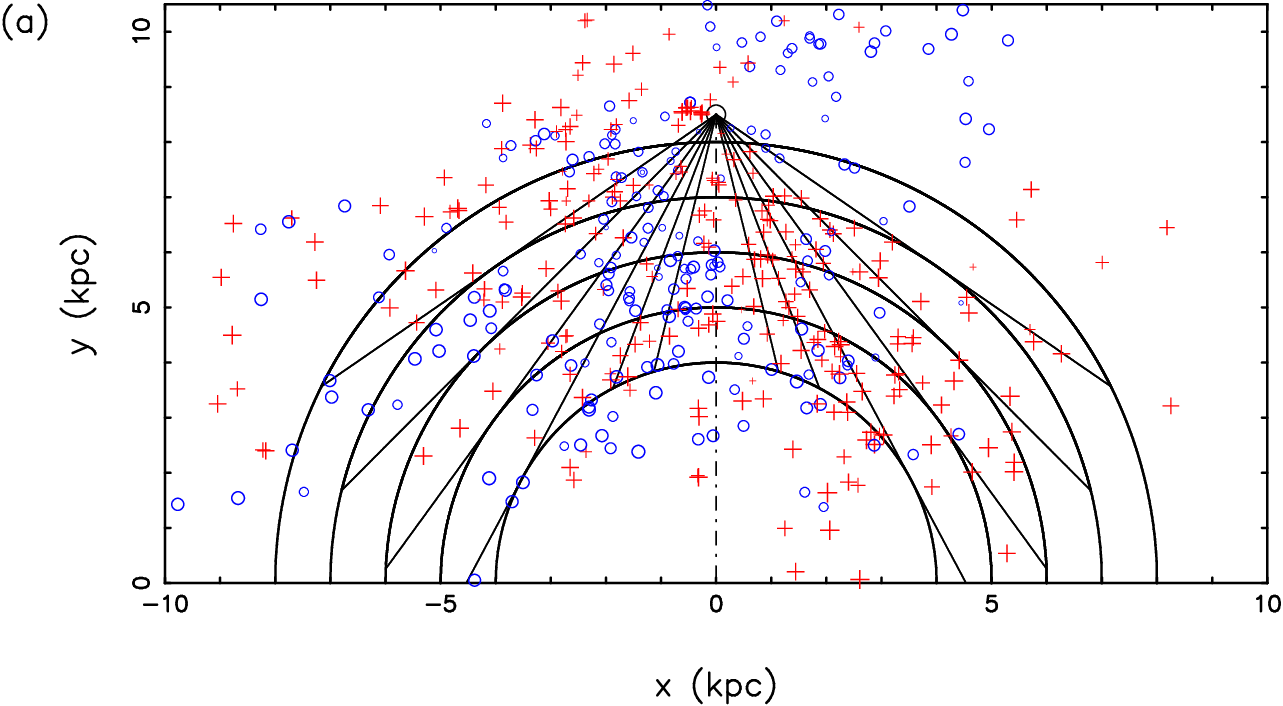}
\bigskip\bigskip\\
\includegraphics[angle=-90,width=12cm]{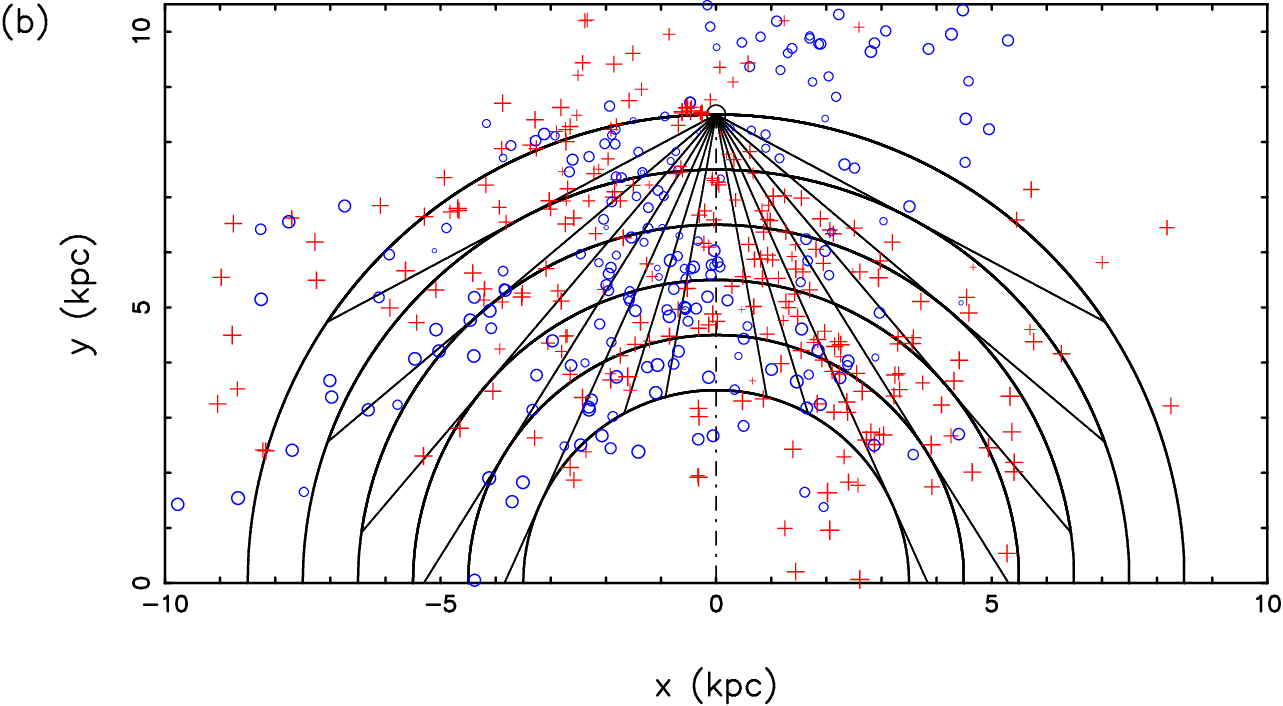}
\caption{\label{Fig.2}
  Grids used in our study overlaid on the face-on distribution of our
  482 low-latitude ($|b|<10\degr$) pulsars. Pulsars with a positive
  (negative) RM are denoted with crosses (circles). ($x$,$y$) are
  Galactocentric cartesian coordinates. The Sun is located at
  ($x=8.5 \ {\rm kpc}, y=0$). In the upper panel, the circles underlying the
  grids are at $r = 4\ {\rm kpc},\ 5\ {\rm kpc},\ 6\ {\rm kpc},\ 7\
  {\rm kpc} {\rm \ and}\ 8\ {\rm kpc}$, while in the lower panel, they
  are at $r = 3.5\ {\rm kpc},\ 4.5\ {\rm kpc},\ 5.5\ {\rm kpc},\ 6.5\
  {\rm kpc}, \ 7.5\ {\rm kpc} {\rm \ and}\ 8.5\ {\rm kpc}$.
}
\end{figure*}

   In order to map out the large-scale component of
   $\overline{B_{||}}$, one needs to divide the Galactic disk into
   regions (boxes) having sizes intermediate between the large scales
   of the regular field and the small scales of the turbulent field
   and containing at least a few pulsars each. In previous studies
   \citep{RL94, Weisberg04, Han06}, this division was based on a
   heliocentric grid defined by circles of constant $d$ and radial
   lines of constant $l$. Such a heliocentric division was justified
   by the spatial distribution of the available pulsars, but it is
   ill-suited to the present work, whose purpose is to test field
   models expressed in terms of Galactic radius, $r$. A much more
   appropriate division here is one based on a hybrid grid defined by
   circles of constant $r$ and lines of constant $l$ (see
   Fig.~\ref{Fig.2}).
To make full use of the pulsar data, we consider two different grids.
In the first grid, the circles are located at $r = 4\ {\rm kpc},\
5\ {\rm kpc},\ 6\ {\rm kpc},\ 7\ {\rm kpc} {\rm \ and}\ 8\ {\rm kpc}$
[i.e., $r=r_i$, with $r_i \equiv i\ {\rm kpc},\ i=4\, ...\, 8$], and
the lines of constant $l$ are the lines emanating from the Sun and
tangent to one of the circles $r=r_i,\ i=2\, ...\, 7$, plus the line Sun-GC
[i.e., $l=l_0,\ l_{\pm i}$, with $l_0 \equiv 0$ and $l_{\pm i} \equiv
\pm {\rm asin} \frac{r_i}{r_\odot},\ i=2\, ...\, 7$]
(see Fig.~\ref{Fig.2}a).
The second grid is defined in an analogous manner with the circles shifted
by 0.5~kpc [i.e., $r=r_i,\ i=3.5,\, 4.5\, ...\, 8.5$,
and accordingly, $l=l_0,\ l_{\pm i},\ i=1.5,\, 2.5\, ...\, 7.5$]
(see Fig.~\ref{Fig.2}b).
To ensure a sufficient number of pulsars per box, some of the boxes
defined by these grids are paired together.
More specifically, the non-outermost boxes along each ring are paired
either with their left or right neighbor along the same ring (thereby
leading to a single double-size box) or with both neighbors separately
(thereby leading to two overlapping boxes).

%___________________  Fig 2  ________________________________

%____________________________________________________________

  Altogether, we have 9 rings, which are centered on $r_i \equiv i\ {\rm kpc}$,
  with $i= 4.5,\, 5.5\, ...\, 7.5$ in the first grid and $i= 4,\, 5\, ...\, 8$
  in the second grid. In the following, the ring centered on $r_i$ (and
  extending between $r_{i-0.5}$ and $r_{i+0.5}$) is referred to as ring $i$.

We retain only the boxes containing at least 5 RMs. 
This minimum number of RMs, which is smaller than generally recommended 
for statistical testing, results from the limited number of pulsars with 
know RMs;
it was chosen as a trade-off between the need to have enough data points 
per box to make use of Eq.~(\ref{Eq_B_par_obs}) and the need to have 
enough boxes on the Galactic plane to capture the spatial variations 
of the large-scale magnetic field.
Even with such a small number of RMs in a given box, it is possible 
to trace the dependences of RM on distance and on DM for the pulsars 
lying in this box \citep[see][]{RL94, Weisberg04, Han06}.

\begin{figure}[!htb]
\centering
\includegraphics[angle=-90,width=7cm]{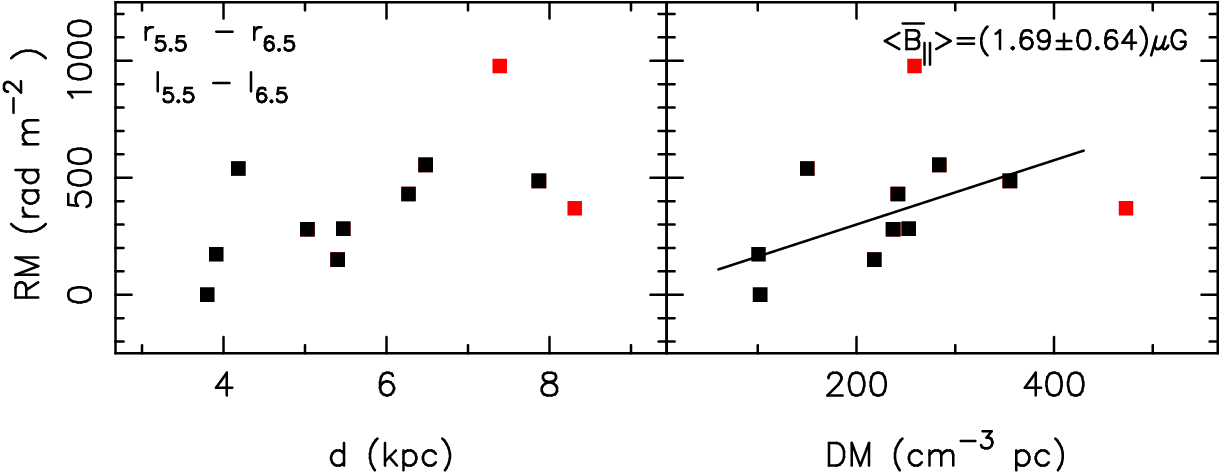}
\bigskip\\
\includegraphics[angle=-90,width=7cm]{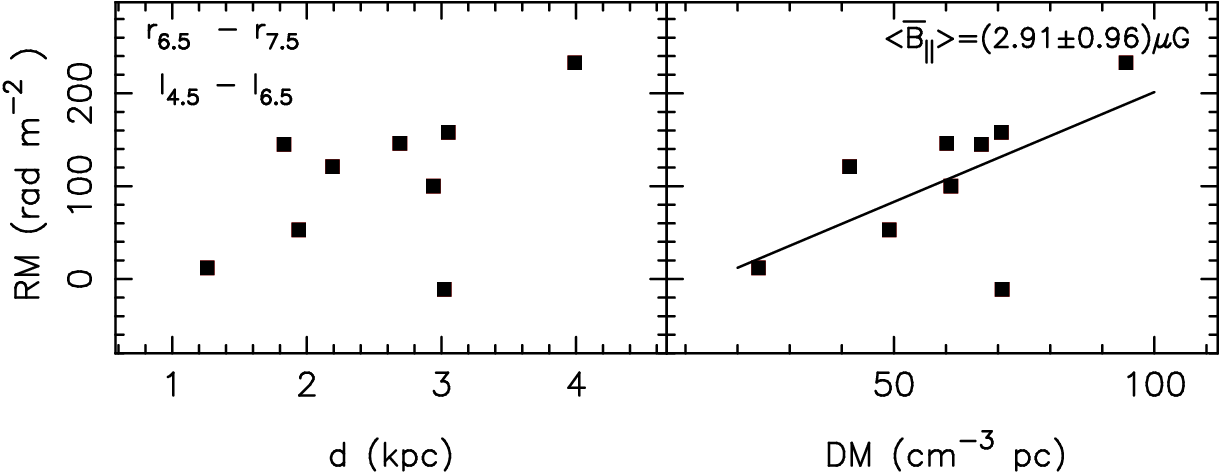}
\caption{\label{Fig.epl}
Plots of RM {\it versus} distance (left panels) and {\it versus} DM 
(right panels) for the pulsars lying in two different boxes.
The first box (top row) is delimited by the circles $r=r_{5.5}$ and
$r=r_{6.5}$ in the radial direction and by the tangential lines 
$l=l_{5.5}$ and $l=l_{6.5}$ in the longitudinal direction.
The second box (bottom row) is delimited by the circles $r=r_{6.5}$ and
$r=r_{7.5}$ and by the tangential lines $l=l_{4.5}$ and $l=l_{6.5}$.
For each box, the best-fit straight line through the points (DM,RM) 
is drawn in the right panel, and the corresponding value of 
$\left< \overline{B_{||}} \right>$ with its statistical uncertainty 
are written in the upper right corner.
The red points represent outliers.
}
\end{figure}

We estimate the average value of $\overline{B_{||}}$, denoted by 
$\left< \overline{B_{||}} \right>$, in each of the retained boxes 
in the following way: we plot the points (DM,RM) of all the pulsars 
in the box, we fit a straight line through the resulting set of data points,
and we take $\left< \overline{B_{||}} \right>$ to be 1.232 times the slope 
of this line (see Eq.~(\ref{Eq_B_par_obs}) and preceding comment).
  To perform the straight-line fit, we resort to a slightly modified
  version of the ordinary least-squares linear regression of $Y$ on
  $X$ described by \citet{Isobe90},
  which is well suited when the dispersion of the data points about
  the linear relation cannot be calculated beforehand.
  The modifications brought to the original scheme are designed to exclude
  the occasional outliers -- such as those arising from H{\sc ii} regions
  \citep{Mitra03}. In practice, we discard all the data points
  whose absolute deviation from the mean RM or DM exceeds three times
  the mean absolute deviation.
To illustrate the procedure, we show two examples in Fig.~\ref{Fig.epl}.

%________________  Fig 3  ___________________________________
%____________________________________________________________

\begin{figure*}[!htb]
\centering
\includegraphics[angle=-90,width=12cm]{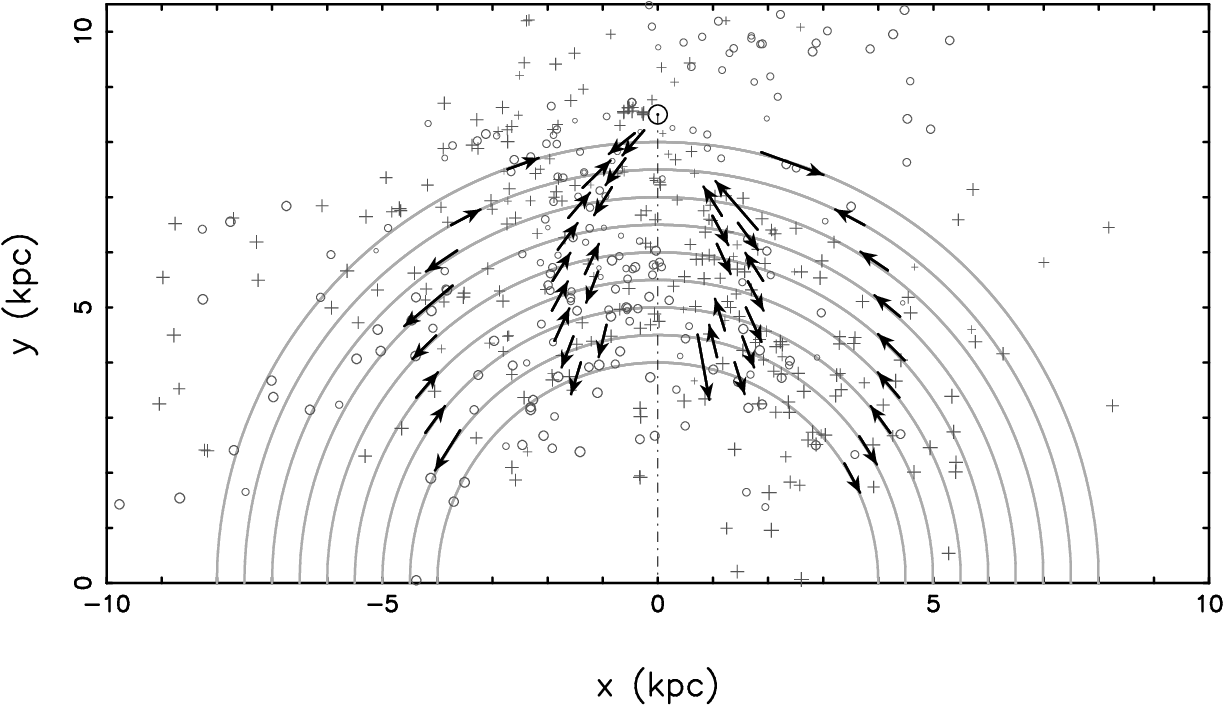}
\caption{\label{Fig.3}
Face-on map showing the average line-of-sight component of the magnetic
field, $\left<\overline{B_{||}}\right>$, obtained in the different boxes
of our two grids.
Each $\left<\overline{B_{||}}\right>$ is plotted in the form of a vector
centered on the box midpoint (point at middle radius, $r=r_i$,
and middle longitude, $l = (l_{\rm min} + l_{\rm max}) /2$),
and oriented along the local line of sight.
}
\end{figure*}

The derived values of $\left< \overline{B_{||}} \right>$ 
in all the boxes of our two grids are mapped in Fig.~\ref{Fig.3}.
For convenience, these values are converted into vectors oriented along
the local line of sight.

%________________  Fig 3  ___________________________________

%____________________________________________________________

As explained above Eq.~(\ref{Eq_B_par_obs}), the turbulent component
of the magnetic field, $\delta {\bf B}$, causes the RMs to scatter
about the mean DM-RM line. This physical scatter due to turbulence
is typically one order of magnitude larger than the observational
scatter due to measurement errors.  The amplitude of the RM scatter
is given by the r.m.s. deviation of the measured RMs from the mean DM-RM
line.  Since the RM scatter is of predominantly turbulent origin,
its amplitude divided by the mean DM in the considered box directly yields 
(to a factor 1.232) an estimate for the r.m.s. value of the turbulent 
component of the line-of-sight field, $\delta \overline{B_{||}}$.
Finally, the r.m.s. value of $\delta \overline{B_{||}}$ divided by 
the square root of the number of pulsars in the box provides an estimate 
for the statistical uncertainty in $\left< \overline{B_{||}} \right>$, 
$\sigma_{||}$, which, again, is predominantly due to turbulence.
The exact expression of $\sigma_{||}$ can be found in \citet{Isobe90}.
With our data, the typical values of $\sigma_{||}$ lie between 
$\sim 0.2\, \mu{\rm G}$ and $1.3 \, \mu{\rm G}$.

\section{Data fitting to the field models}

Once we have obtained a set of observational values of 
$\left< \overline{B_{||}} \right>$ together with their statistical
uncertainties (or error bars), we can put the three theoretical
field models presented in Sect.~2 to the test. 
As in all other studies based on RMs, we proceed on the notion that
the large-scale interstellar magnetic field may be identified with 
its $n_{\rm e}$-weighted average value (denoted with an overbar).
Implicit here is the assumption that fluctuations in magnetic field 
strength and in free-electron density are statistically uncorrelated.
In reality, this assumption is certainly not strictly satisfied in the ISM
\citep[e.g.,][]{Beck03}, and this will cause our results to be somewhat biased.

With this caveat in mind, we now describe the overall procedure. 
For each model, we use all our observational values
of $\left< \overline{B_{||}} \right>$ to derive the best-fit
parameters of the model. We then examine whether the best fit is
consistent with the pulsar data, i.e., whether the theoretical
line-of-sight fields predicted by it fall within the error bars
of the observational $\left< \overline{B_{||}} \right>$
(in a statistical sense).
If we find that the best fit is not consistent with the data,
we may conclude that the considered model must be rejected.  
If, on the other hand, the best fit is found consistent with the data, 
we may conclude that the model is acceptable; we then determine 
the extent of the so-called "consistency domain", i.e., 
the parameter domain around the best fit within which solutions 
are consistent with the data.

It is important to realize that the concept of acceptability
differs from the concept of detectability.
A given field model, say, the ring model, is acceptable
only if it is not ruled out by the available pulsar data.
This does not necessarily imply that the Galactic magnetic field
is really of the ring type, nor that a ring field has truly
been detected. Detection of a ring field requires not only that
the ring model be acceptable, but also that the zero-field solution
do not belong to the consistency domain.

Let us now discuss more specifically what exact criterion should be 
used to test consistency with the pulsar data for a given field model.
Each of the three models is characterized by a number of independent
free functions of Galactic radius [$\left<B_\theta\right>(r)$ 
in the ring model; $\left<B_r\right>(r)$ and $\left<B_\theta\right>(r)$ 
in the axisymmetric model; $b_r(r)$, $b_{\theta}(r)$ and $\phi(r)$
in the bisymmetric model], corresponding to the same number of
independent free parameters in every ring $i$ [denoted by 
$\left<B_\theta\right>_i$ in the ring model; $\left<B_r\right>_i$ 
and $\left<B_\theta\right>_i$ in the axisymmetric model; 
$b_{r,i}$, $b_{\theta,i}$ and $\phi_i$ in the bisymmetric model].
Therefore, the 9 different rings may be analyzed separately.

For any one of the three field models, consider a given ring $i$ 
and suppose that this ring contains $n_i$ boxes.
For every box $j$, we have derived an observational value of
the average line-of-sight field $\left<\overline{B_{||}}\right>$,
denoted by $\left<\overline{B_{||}}\right>_{ij}$, together with
its statistical uncertainty, denoted by $(\sigma_{||})_{ij}$.
Besides, we can calculate a theoretical expression of the large-scale 
line-of-sight field $\left<B_{||}\right>$, denoted by 
$\left<B_{||}\right>_{ij}$, in terms of the free parameters of ring $i$.
%%for the model under test.
The best-fit values of these parameters are obtained by minimizing
   \begin{equation}
	 \chi^2 = \sum^{n_i}_{j=1}
	 \left( \frac{ \left<\overline{B_{||}}\right>_{ij} -
				   \left<B_{||}\right>_{ij} }
				 { (\sigma_{||})_{ij} }
	 \right)^2 \ .
	 \label{Eq_chi2_par}
   \end{equation}

The best fit of ring $i$ can be considered consistent with the pulsar data 
if, on average over ring $i$, the theoretical best-fit 
$\left<B_{||}\right>_{ij}$ do not differ from the observational 
$\left<\overline{B_{||}}\right>_{ij}$
by more than the associated uncertainties $(\sigma_\theta)_{ij}$.
In mathematical terms, this condition for consistency can be expressed as 
$\chi^2 \le n_i $. 
However, when the number of data points, $n_i$, is not much greater than 
the number of free parameters, $\nu$ [$\nu=1$ for the ring model;
$\nu=2$ for the axisymmetric model; $\nu=3$ for the bisymmetric model], 
consistency with the data should be tested with the more exact criterion
   \begin{equation}
	 \chi^2 \le n_i - \nu \ ,
	 \label{Eq_consist_par}
   \end{equation}
where $n_i -\nu$ is the number of degrees of freedom, i.e., 
the number of data points that cannot automatically be placed on a curve
with $\nu$ adjustable parameters.
Eq.~(\ref{Eq_consist_par}) provides a rule of thumb for a reasonably
good fit \citep[see Sect.~15.1 in][]{ptvf92}.
If $\chi^2 \gg n_i -\nu$, the best-fit curve misses too many data points 
to be believable.

It is possible to obtain a more rigorous (and, at the same time, 
more flexible) criterion for consistency.
Suppose, for the sake of argument, that the model we are testing is correct. 
If the data points $\left<\overline{B_{||}}\right>_{ij}$ of ring $i$ 
follow a Gaussian distribution, $\chi^2$ has a chi-square distribution 
with $n_i -\nu$ degrees of freedom. 
One can then calculate the {\it a priori} probability,
$P [\chi^2>\chi^2_{\rm crit}]$, that the $\chi^2$ obtained 
for a particular set of data points exceeds some critical value 
$\chi^2_{\rm crit}$.
Conversely, one can calculate the critical $\chi^2_{\rm crit}$
for which $P [\chi^2>\chi^2_{\rm crit}]$ equals some imposed 
probability $P_0$.
For instance, if the model is correct, it is unlikely (only 10\% chance) 
that $\chi^2 > \chi^2_{\rm crit} (P_0=0.1)$. Turning the statement around, 
if we find $\chi^2 > \chi^2_{\rm crit} (P_0=0.1)$, it is unlikely that 
the model is correct -- we will say that the model is inconsistent 
with the data.
This reasoning directly leads to the following consistency condition:
   \begin{equation}
         \chi^2 \le \chi^2_{\rm crit} (P_0) \ .
	 \label{Eq_consist_prob}
   \end{equation}
Here, we will adopt $P_0=0.1$ as our default value, 
but we will also discuss the results obtained for $P_0=0.05$.
In practice, the values of $P [\chi^2>\chi^2_{\rm crit}]$
for given $n_i -\nu$ and $\chi^2_{\rm crit}$ are tabulated in various
textbooks \citep[e.g.,][]{Yamane64}. The tables can also be used
to determine $\chi^2_{\rm crit} (P_0)$ for given $n_i -\nu$ and $P_0$.
For reference, the values of $\chi^2_{\rm crit} (P_0)$ 
for $n_i-\nu=1,\, 2,\, 3,\, 4,\, 5$ and for $P_0=0.05,\, 0.1,\, 0.2$ 
are listed in Table~\ref{Table_Crit}.
An important point emerging from Table~\ref{Table_Crit} is that 
$\chi^2_{\rm crit} (P_0=0.1) > n_i -\nu$, 
so that Eq.~(\ref{Eq_consist_prob}) with $P_0=0.1$ will always be 
easier to satisfy than Eq.~(\ref{Eq_consist_par}) .

\begin{table}
\caption{Critical values of $\chi^2$ for 3 probability levels}
\label{Table_Crit}
\centering
\begin{tabular}{c|c c c }
\hline\hline
\noalign{\smallskip}
%%\backslashbox{$n_i -\nu$}{$P_0$} & 0.05 & 0.1 & 0.2\\
$n_i -\nu$ & $\chi^2_{\rm crit} (P_0=0.05)$ & 
$\chi^2_{\rm crit} (P_0=0.1)$ & $\chi^2_{\rm crit} (P_0=0.2)$ \\
\hline
\noalign{\smallskip}
1 & 3.841 & 2.706 & 1.642 \\
2 & 5.991 & 4.605 & 3.219 \\
3 & 7.815 & 6.251 & 4.642 \\
4 & 9.488 & 7.779 & 5.989 \\
5 & 11.070 & 9.236 & 7.289 \\
\noalign{\smallskip}
\hline
\end{tabular}
\end{table}

In the next three subsections, we present the results obtained 
with the rule of thumb (Eq.~(\ref{Eq_consist_par}))
and with the more rigorous consistency condition 
(Eq.~(\ref{Eq_consist_prob})), for the three field models.

\subsection {Ring model}

In the ring model, $\left<B_r\right>$ vanishes and $\left<B_\theta\right>$
is constant along circles. Hence, there are 9 free parameters: 
$\left<B_\theta\right>_i$, the large-scale azimuthal fields 
in the 9 rings $i= 4,\, 4.5,\, 5\, ...\, 8$.\footnote{\label{footnote1}
As a reminder, ring $i$ is centered on $r_i \equiv i\ {\rm kpc}$ and 
extends between $(i-0.5)\ {\rm kpc}$ and $(i+0.5)\ {\rm kpc}$.}
Since all the free parameters are independent, the 9 rings can be treated
separately.

For every ring $i$, the large-scale line-of-sight field in any box $j$ 
is simply the projection of $\left<B_\theta\right>_i$ onto the line of
sight (see Eq.~(\ref{Eq_B_para}) with $\left<B_r\right> = 0$):
   \begin{equation}
	\left<{B_{||}}\right>_{ij} =
		\left<{B_\theta}\right>_{i}\ \cos \alpha_{ij} \ ,
   \label{Eq_ring_Bpij}
   \end{equation}
where $\alpha_{ij}$ is the angle between the azimuthal direction and
the direction to the Sun at the midpoint\footnote{As in Fig.~\ref{Fig.3},
the midpoint of a box is defined as the point at middle radius, $r=r_i$, 
and middle longitude, $l = (l_{\rm min} + l_{\rm max}) /2$.} of box $j$ 
(see Fig.~\ref{Fig.1}). 
The best-fit value of $\left<B_\theta\right>_i$ is obtained 
by minimizing $\chi^2$ (given by Eq.~(\ref{Eq_chi2_par})).
In terms of $\left<\overline{B_\theta}\right>_{ij} =
\left<\overline{B_{||}}\right>_{ij} / \cos \alpha_{ij}$,
the observational value of the average azimuthal field in box $j$,
and $(\sigma_\theta)_{ij} = (\sigma_{||})_{ij} / \cos \alpha_{ij}$,
the associated uncertainty, the minimization procedure turns out to be
equivalent to taking an uncertainty-weighted average of the different 
$\left<\overline{B_\theta}\right>_{ij}$ along ring $i$:
   \begin{equation}
     \left<B_\theta\right>_i \ = \
     \frac{\displaystyle
           \sum^{n_i}_{j=1} \frac{\left<\overline{B_\theta}\right>_{ij}}
                                 {(\sigma_\theta)_{ij}^2}
          }
          {\displaystyle
           \sum^{n_i}_{j=1} \frac{1}
                                 {(\sigma_\theta)_{ij}^2}
          }
     \ .
     \label{Eq_ring_i}
   \end{equation}

\begin{figure*}[!htb]
\centering
\includegraphics[angle=-90,width=10cm]{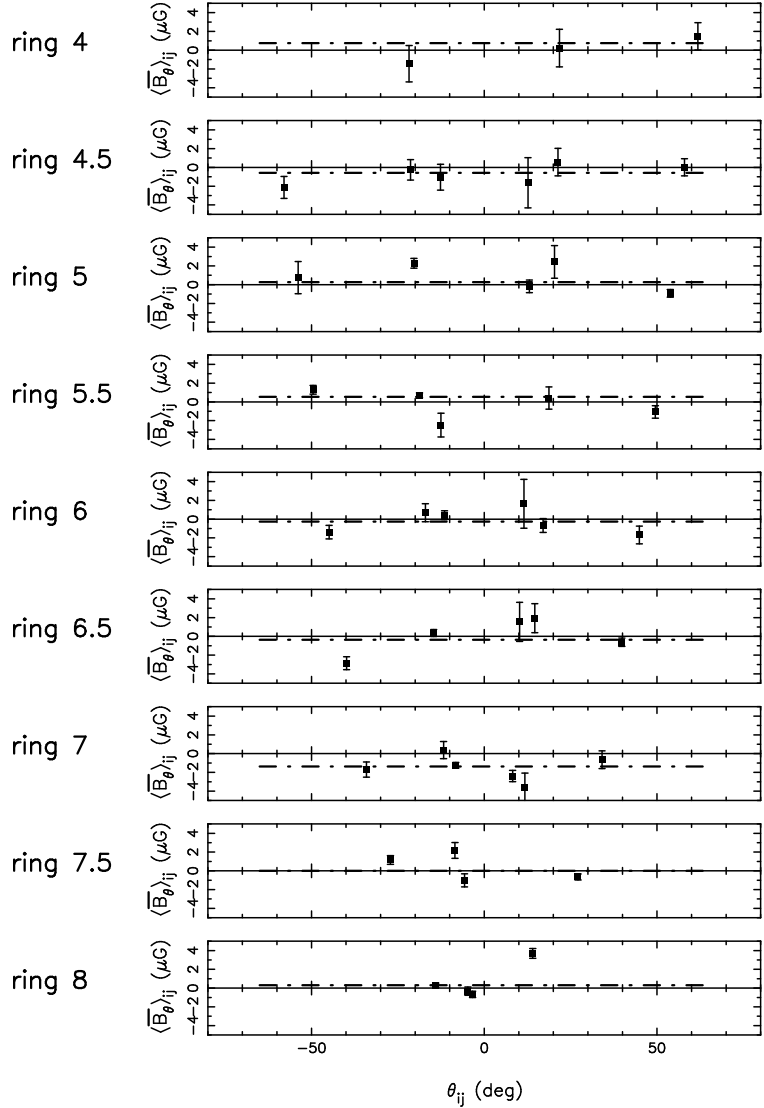}
\caption{\label{Fig_ring_B_theta}
Observational values of the average azimuthal fields,
$\left<\overline{B_\theta}\right>_{ij}$, {\it versus}
Galactic azimuthal angles, $\theta_{ij}$, in all the boxes $j$
of the 9 successive rings $i$, for the ring model.
The uncertainties $(\sigma_\theta)_{ij}$ in the field values
are plotted as standard error bars.
For each ring $i$, the best-fit value of the model parameter
$\left<B_\theta\right>_i$ is indicated by the horizontal
dot-dashed line.
}
\end{figure*}

The values of $\left<\overline{B_\theta}\right>_{ij}$
and their uncertainties $(\sigma_\theta)_{ij}$ in the $n_i$ boxes $j$
of the 9 rings $i$ are plotted in Fig.~{\ref{Fig_ring_B_theta}},
at the Galactic azimuthal angles of the box midpoints, $\theta_{ij}$.
For comparison, the best-fit values of $\left<B_\theta\right>_i$
in the 9 rings are indicated by horizontal lines spanning the entire
azimuthal range.

%_________________ Fig 4 _________________________________

%____________________________________________________________

\begin{figure*}[!htb]
\centering
\includegraphics[angle=-90,width=8cm]{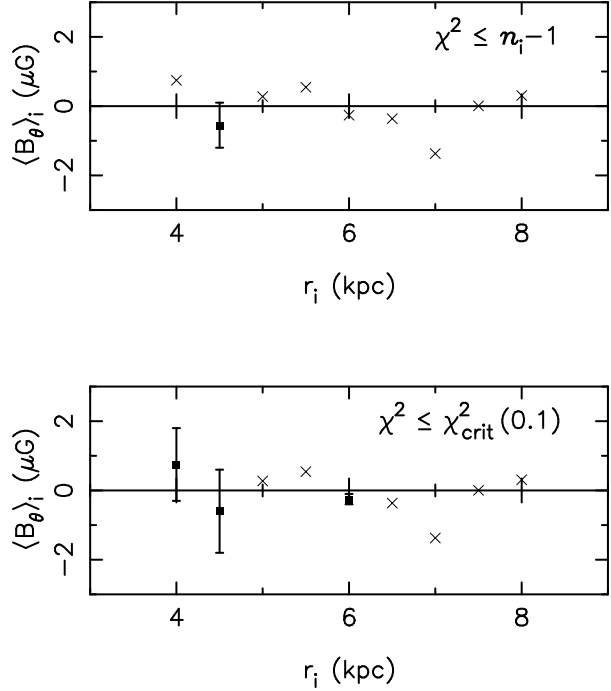}
\caption{\label{Fig_ring_B_r}
Best-fit values of the model parameters, $\left<B_\theta\right>_i$,
{\it versus} ring middle radii, $r_i$, for the ring model.
The acceptable values (those consistent with the pulsar data) are shown
with their consistency ranges, while the unacceptable values are
indicated with crosses.
The results obtained with the rule-of-thumb consistency condition,
$\chi^2  \le n_i-1$, are plotted in the upper panel.
Those obtained with the more rigorous consistency condition,
$\chi^2 \le \chi^2_{\rm crit} (0.1)$,
are plotted in the lower panel.
}
\end{figure*}

In only one ring ($i=4.5$) does the best-fit value of
$\left<B_\theta\right>_i$ satisfy the rule-of-thumb 
consistency condition, $\chi^2 \le n_i - 1$
(Eq.~(\ref{Eq_consist_par}) with $\nu=1$).
For this ring, we compute the consistency range of
$\left<B_\theta\right>_i$, which contains all the values of
$\left<B_\theta\right>_i$ for which $\chi^2 \le n_i - 1$.
The best-fit value of $\left<B_\theta\right>_i$ and its consistency
range in the sole "good-fit" ring are plotted against $r_i$,
in the upper panel of Fig.~{\ref{Fig_ring_B_r}}.
For the other 8 rings, the (inconsistent) best-fit values of
$\left<B_\theta\right>_i$ are plotted with crosses.
Clearly, these 8 rings do not admit any ring magnetic field 
consistent with the data.
As an immediate consequence, the ring model must be rejected.

In order to gain some feel for how far the ring model is from being
able to reproduce the pulsar data, let us, in thought, extend the
error bars of all the observational
$\left<\overline{B_{||}}\right>_{ij}$ by a factor of 2 and look into
the impact of this extension on our results. With twice the original
error bars, the $\chi^2$ parameter would be smaller by a factor of 4,
so that, in terms of the original $\chi^2$, the rule-of-thumb
consistency condition would become $\chi^2  \le 4 (n_i-1)$.  
As it turns out, this less stringent consistency condition would be 
fulfilled in  5 rings ($i=4, \, 4.5, \, 5.5, \, 6, \, 7$)
out of 9. Thus, with twice the original error bars, the ring model 
would remain unacceptable.

If we now resort to the more rigorous consistency condition,
$\chi^2 \le \chi^2_{\rm crit} (0.1)$
(Eq.~(\ref{Eq_consist_prob}) with $P_0=0.1$),
to test the ring model, we find that 3 rings ($i=4, \, 4.5, \, 6$) 
have their best-fit $\left<B_\theta\right>_i$ consistent with the data;
their consistency ranges are drawn in the lower panel of 
Fig.~{\ref{Fig_ring_B_r}}.
For the other 6 rings, the (inconsistent) best-fit $\left<B_\theta\right>_i$ 
are again plotted with crosses.
With $P_0=0.05$, 4 rings ($i=4, \, 4.5, \, 6,\, 7$) would be deemed
consistent with the data, but the other 5 rings would still fail
the consistency test. 

These results confirm our conclusion that the ring model must be rejected.

%_________________  Fig 5____________________________________
%
%____________________________________________________________

\subsection {Axisymmetric model}

In the axisymmetric model, $\left<B_r\right>$ and $\left<B_{\theta}\right>$
are both constant along circles. Hence, there are 18 free parameters: 
$\left<B_r\right>_i$ and $\left<B_{\theta}\right>_i$,
the large-scale radial and azimuthal fields
in the 9 rings $i= 4,\, 4.5,\, 5\, ...\, 8$.

For every ring $i$, the large-scale line-of-sight field in any box $j$ 
can be written as a linear combination of the two parameters
$\left<B_r\right>_i$ and $\left<B_{\theta}\right>_i$:
   \begin{equation}
	 \left<B_{||}\right>_{ij}
	 = \left<B_r\right>_i \, \sin \alpha_{ij}
	 + \left<B_{\theta}\right>_i \, \cos \alpha_{ij}
	 \label{Eq_ASS_ij}
   \end{equation}
(see Eq.~(\ref{Eq_B_para})).
Again the best-fit values of $\left<B_r\right>_i$ and 
$\left<B_{\theta}\right>_i$ are obtained by minimizing $\chi^2$ 
(Eq.~(\ref{Eq_chi2_par})).

\begin{figure*}[!htb]
\centering
\includegraphics[angle=-90,width=12cm]{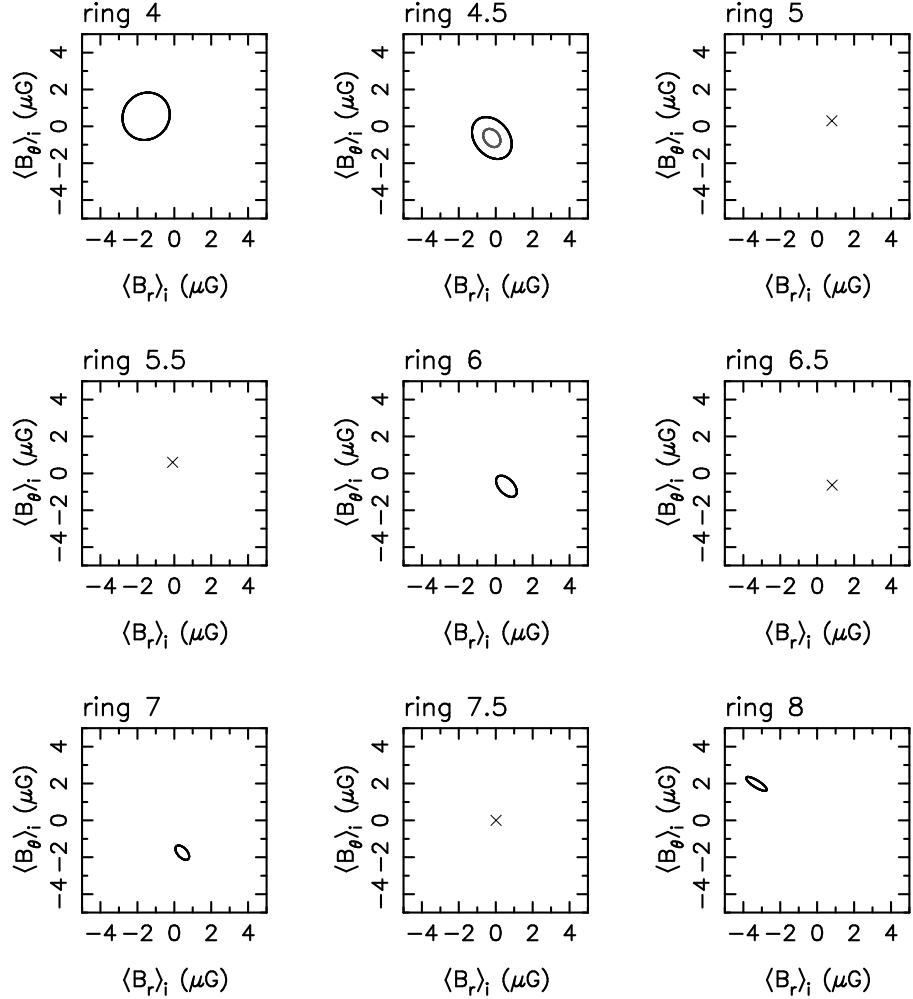}
\caption{\label{Fig_ASS_B_B}
Consistency domains in the parameter planes
$(\left<B_r\right>_i,\left<B_{\theta}\right>_i)$ of the 9 different rings,
for the axisymmetric model.
The grey contour lines define the consistency domains obtained 
with the rule-of-thumb consistency condition, $\chi^2 \le n_i-2$. 
The black contour lines define those obtained with the more rigorous
consistency condition, $\chi^2 \le \chi^2_{\rm crit} (0.1)$.
The crosses mark the locations of the unacceptable best fits
(those inconsistent with the pulsar data according to both criteria).
}
\end{figure*}

Here, we find that 1 ring ($i= 4.5$) has its best fit
consistent with the data, according to the rule-of-thumb consistency 
condition, $\chi^2 \le n_i - 2$ (Eq.~(\ref{Eq_consist_par}) with $\nu=2$).
Its consistency domain in the parameter plane
$(\left<B_r\right>_i,\left<B_{\theta}\right>_i)$ 
is the area delimited by the ellipse $\chi^2 = n_i -2$
(grey contour line in the relevant panel of Fig.~\ref{Fig_ASS_B_B}).
As none of the other 8 rings can be properly fit with an axisymmetric
magnetic field, the axisymmetric model must be rejected.

With twice the original error bars on the observational
$\left<\overline{B_{||}}\right>_{ij}$, the rule-of-thumb consistency 
condition would become $\chi^2 \le 4 (n_i-2)$ (in terms of the original 
$\chi^2$).
This less stringent consistency condition would be satisfied in 5 rings   
($i=4,\, 4.5,\, 6,\, 7,\, 8$), but still not in the other 4 rings.
Therefore, the axisymmetric model would remain unacceptable.

According to the more rigorous consistency condition,
$\chi^2 \le \chi^2_{\rm crit} (0.1)$ (Eq.~(\ref{Eq_consist_prob}) 
with $P_0=0.1$), 5 rings ($i=4,\, 4.5,\, 6,\, 7,\, 8$) have their best fits 
consistent with the data.
Their consistency domains in the parameter planes
$(\left<B_r\right>_i,\left<B_{\theta}\right>_i)$ are the elliptical
areas enclosed by the curves $\chi^2 = \chi^2_{\rm crit} (0.1)$
(black contour lines in the relevant panels of Fig.~\ref{Fig_ASS_B_B}).
For the other 4 rings, the (inconsistent) best-fit pairs
$(\left<B_r\right>_i,\left<B_{\theta}\right>_i)$
are indicated with crosses. 
Relaxing the probability level to $P_0=0.05$ would not raise
the number of acceptable rings above 5.

From all the above, we conclude that the axisymmetric model must 
be rejected.

%_________________ Fig 7 ____________________________________

%______________________________________________________

\subsection {Bisymmetric model}

In the bisymmetric model, $\left< B_r \right>$ and $\left< B_\theta \right>$
vary sinusoidally along circles in the manner described by
Eqs.~(\ref{Eq_BSS_Br}) and (\ref{Eq_BSS_Btheta}).
Hence, there are 27 free parameters:
$b_{r,i}$, $b_{\theta,i}$ and ${\phi}_i$, the maximum amplitudes and
the azimuthal phases in the 9 rings $i= 4,\, 4.5,\, 5\, ...\, 8$.

For every ring $i$, the large-scale line-of-sight field in any box $j$ 
follows from Eq.~(\ref{Eq_B_para}) together with
Eqs.~(\ref{Eq_BSS_Br})--(\ref{Eq_BSS_Btheta}):
	 \begin{equation}
	   \left<{B_{||}} \right>_{ij} =
	   b_{r,i} \, \sin (\theta_{ij}-\phi_{i})\, \sin \alpha_{ij} +
	   b_{\theta,i} \, \sin(\theta_{ij}-\phi_{i}) \, \cos \alpha_{ij} \ ,
	   \label{Eq_BSS_ij}
	 \end{equation}
where the angles $\theta_{ij}$ and $\alpha_{ij}$ (see Fig.~\ref{Fig.1})
refer to the midpoint of box $j$.
Similarly to the previous models, the best-fit values of
the three parameters $b_{r,i}$, $b_{\theta,i}$ and ${\phi}_i$
are obtained through a minimization of $\chi^2$ (Eq.~(\ref{Eq_chi2_par})).

% It is important to note that, in the outermost ring ($i=8$), 
% the number of data points ($n_i=3$) is exactly equal to 
% the number of adjustable parameters ($\nu=3$).
% This means that we will be able to find a curve that passes exactly
% through the three data points, but with zero degree of freedom left
% ($n_i -\nu = 0$), we will have no way of testing the goodness of fit.
% For this reason, we exclude the outermost ring from our discussion 
% of the model's acceptability.

\begin{figure*}[!htb]
\centering
\includegraphics[angle=-90,width=12cm]{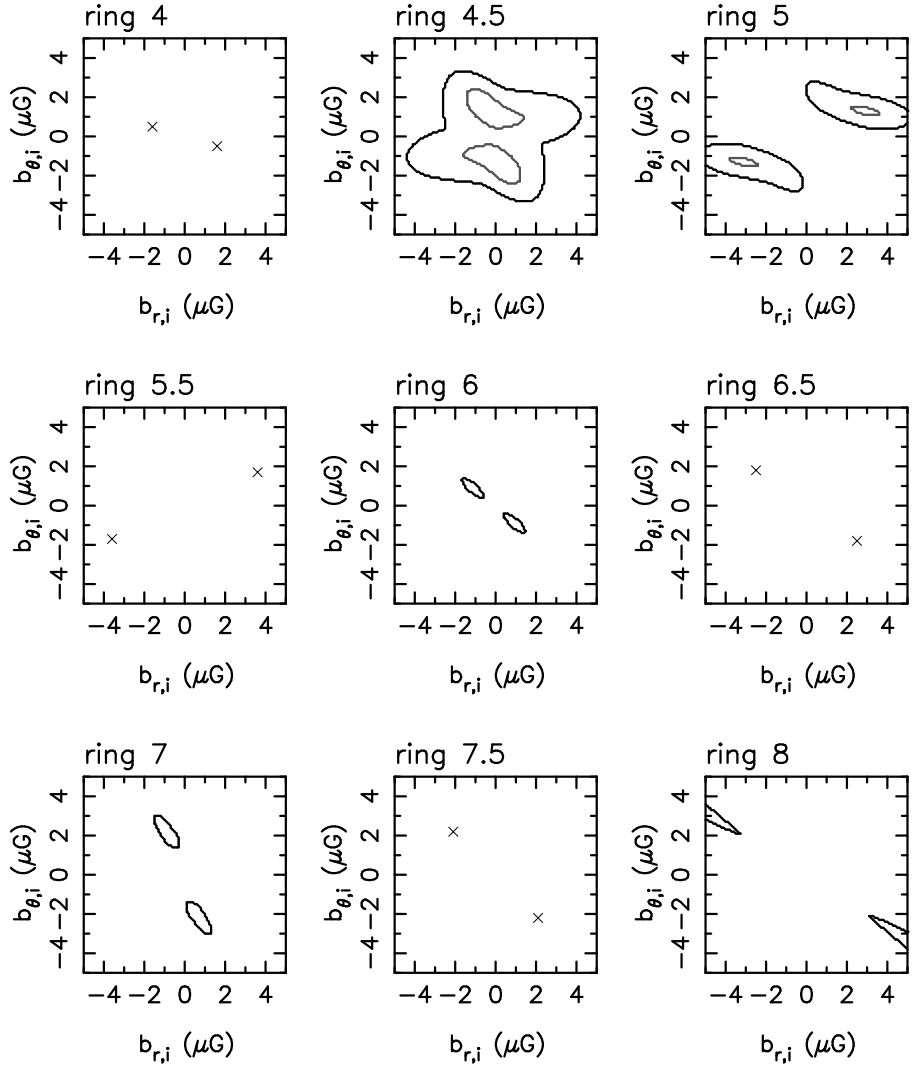}
\caption{\label{Fig_BSS_b_b}
Projections of the 3D consistency domains of the 9 different rings
on their parameter planes $(b_{r,i},b_{\theta,i})$, for the bisymmetric model.
The grey contour lines correspond to the consistency domains obtained 
with the rule-of-thumb consistency condition, $\chi^2 \le n_i-3$.
The black contour lines correspond to those obtained with the more
rigorous consistency condition, $\chi^2 \le \chi^2_{\rm crit} (0.1)$.
The crosses mark the locations of the unacceptable best fits
(those inconsistent with the pulsar data according to both criteria). 
%The untestable best fit of ring~8 is not shown because it falls 
%  outside the frame.
}
 \end{figure*}

According to the rule-of-thumb consistency condition,
$\chi^2 \le n_i - 3$ (Eq.~(\ref{Eq_consist_par}) with $\nu=3$), 
2 rings ($i=4.5,\, 5$) have their best fits consistent with the data. 
Their consistency domains in the parameter spaces
$(b_{r,i},b_{\theta,i},{\phi}_i)$ are the volumes bounded 
by the surfaces $\chi^2 = n_i - 3$.
Displayed in Fig.~\ref{Fig_BSS_b_b} are the projections of these
consistency domains on the parameter planes $(b_{r,i},b_{\theta,i})$
(grey contour lines).
Since the other 7 rings fail the consistency test, the bisymmetric model
must be rejected.

With twice the original error bars on the observational
$\left<\overline{B_{||}}\right>_{ij}$, the rule-of-thumb consistency
condition would become $\chi^2 \le 4 (n_i-3)$, which would be satisfied
in 7 rings ($i=4,\, 4.5,\, 5, \, 5.5,\, 6,\, 7, \, 8$) out of 9.
The bisymmetric model would then be nearly acceptable.

According to the more rigorous consistency condition,
$\chi^2 \le \chi^2_{\rm crit} (0.1)$ (Eq.~(\ref{Eq_consist_prob}) 
with $P_0=0.1$), 5 rings ($i=4.5,\, 5, \, 6,\, 7, \, 8$) have their best fits
consistent with the data.
Their consistency domains, bounded by the surfaces 
$\chi^2 = \chi^2_{\rm crit} (0.1)$, are also shown in projection
on the parameter planes $(b_{r,i},b_{\theta,i})$ in Fig.~\ref{Fig_BSS_b_b}
(black contour lines).
For the other  4 rings, the (inconsistent) best-fit pairs 
$(b_{r,i},b_{\theta,i})$ are indicated with crosses.
Relaxing the probability level to $P_0=0.05$ would raise the number 
of acceptable rings to 7 ($i=4,\, 4.5,\, 5, \, 5.5,\, 6,\, 7,\, 8$),
which would render the bisymmetric model nearly globally acceptable.

Altogether, the bisymmetric model must be rejected, though its rejection 
is slightly less severe than for the axisymmetric model.

%___________________ Fig 8 __________________________________

%____________________________________________________________

\section {Summary and conclusions}

In this paper, we examined the three most common theoretical models for
the large-scale magnetic field in the Galactic disk and confronted each
of these models with the pulsar data.
For each model, we derived the best-fit parameters, through $\chi^2$
minimization, in the 9 Galactocentric rings defined in Fig.~\ref{Fig.2},
and we delineated the parameter domains around the best fits (referred to 
as the consistency domains) wherein the predicted fields are consistent 
with the pulsar data.

Compared to existing studies of the kind, we did not attempt to settle
the long-standing (and possibly ill-posed) question of whether
the Galactic magnetic field is axisymmetric or bisymmetric.
%%Nor did we try to find out which of these two types of azimuthal symmetry
%%is favored by the pulsar data.
Our sole purpose was to determine whether each of the three basic field
models, taken separately, is compatible with the available pulsar data
or not.
In this regard, we note that many previous studies did find
a preference for one of the field models, but omitted to put their
preferred model through the crucial "goodness-of-fit" test, which checks
whether the model can indeed reproduce the data within the error bars.

Here, we tested the three field models on the basis of two different 
criteria: first, a standard rule of thumb for a reasonably good fit 
(Eq.~(\ref{Eq_consist_par})), and second, a more rigorous consistency 
condition for a chi-square distribution of $\chi^2$ 
(Eq.~(\ref{Eq_consist_prob}), with the imposed probability 
set to $P_0=0.1$).
These two criteria were successively applied to all the rings separately, 
such that the best fit of ring $i$ was deemed consistent with the pulsar data
if the associated value of $\chi^2$, $\chi^2_{\rm min}$, 
was less than $n_i - \nu$ (first criterion) or less than 
$\chi^2_{\rm crit} (P_0=0.1)$ (second criterion).
A model could then be considered globally acceptable if all the rings 
had their best fits consistent with the data.

The results obtained for the three field models, with both criteria,
are summarized in Table~\ref{Table_summary1}. 
All the rings are listed with their labels, $i$ 
(see footnote~\ref{footnote1}), and their numbers of boxes
(or numbers of data points), $n_i$, from which it is straightforward 
to deduce the numbers of degrees of freedom, $n_i -\nu$ (for the first 
criterion), and the critical $\chi^2_{\rm crit} (P_0=0.1)$ 
(for the second criterion; see Table~\ref{Table_Crit}).
Also given for all the rings are the minimum values of $\chi^2$, 
$\chi^2_{\rm min}$, i.e., the values associated with the best fits, 
as well as the results of both consistency tests (satisfaction of a test 
is indicated with an asterisk), for the three field models.

\begin{table*}[!htb]
\begin{minipage}[t]{\columnwidth}
\caption{Summary of the results obtained for the three field models$\ ^a$
}
\label{Table_summary1}
\centering
\renewcommand{\footnoterule}{}  % to avoid a line before footnotes
\begin{tabular}{l l|r c c|r c c|r c c}
\hline\hline
\noalign{\smallskip}
& & \multicolumn{3}{c|}{Ring model \ ($\nu=1$)} 
& \multicolumn{3}{c|}{Axisymmetric model \ ($\nu=2$)} 
& \multicolumn{3}{c}{Bisymmetric model \ ($\nu=3$)} \\
\noalign{\smallskip}
%%\cline{3-5} \cline{7-9} \cline{11-13} \\
\hline
\noalign{\smallskip}
$i$ & $n_i$ & $\chi^2_{\rm min}$ 
& $\!\!\left[ \chi^2_{\rm min} \le n_i - \nu \right]\!\!$
& $\left[ \chi^2_{\rm min} \le \chi^2_{\rm crit} (0.1) \right]$
& $\chi^2_{\rm min}$ 
& $\!\!\left[ \chi^2_{\rm min} \le n_i - \nu \right]\!\!$
& $\left[ \chi^2_{\rm min} \le \chi^2_{\rm crit} (0.1) \right]$
& $\chi^2_{\rm min}$ 
& $\!\!\left[ \chi^2_{\rm min} \le n_i - \nu \right]\!\!$
& $\left[ \chi^2_{\rm min} \le \chi^2_{\rm crit} (0.1) \right]$ \\
\noalign{\smallskip}
\hline
\noalign{\smallskip}
4   & 4 & 5.03 &        & $\ast$ &  2.91 &        & $\ast$ & 3.08 &        & \\
4.5 & 6 & 3.17 & $\ast$ & $\ast$ &  3.11 & $\ast$ & $\ast$ & 1.75 & $\ast$ & $\ast$ \\
5   & 5 & 25.37&        &        & 18.65 &        &        & 1.79 & $\ast$ & $\ast$ \\
5.5 & 5 & 14.52&        &        & 14.43 &        &        & 4.95 &        & \\
6   & 6 & 8.84 &        & $\ast$ &  5.06 &        & $\ast$ & 5.48 &        & $\ast$ \\
6.5 & 5 & 20.29&        &        & 17.93 &        &        &13.55 &        & \\
7   & 6 & 9.57 &        &        &  5.81 &        & $\ast$ & 3.37 &        & $\ast$ \\
7.5 & 4 & 19.31&        &        & 19.30 &        &        & 7.44 &        & \\
8   & 4 & 52.53&        &        &  3.04 &        & $\ast$ & 0.82 &        & $\ast$ \\
\noalign{\smallskip}
\hline
\end{tabular}
\parbox{14cm}{
\begin{flushleft}
	\tiny{$^a\ $ When a value of $\chi^2_{\rm min}$ satisfies the first or second 
        consistency condition, an asterisk is plotted in the corresponding column.}
\end{flushleft}
}
\end{minipage}
\end{table*}

We found that none of the three field models is acceptable, 
in the sense that none of them can be brought into full agreement 
with the pulsar data.
According to the standard rule-of-thumb consistency condition
(Eq.~(\ref{Eq_consist_par})), all three models must be strongly rejected,
as the ring and axisymmetric models fail to provide a good fit 
(consistent with the data) in all the rings save one, 
while the bisymmetric model fails in all the rings save two.
If the error bars of the observational line-of-sight fields
were enlarged by a factor of 2, the bisymmetric model would not be 
too far from acceptable (7 good-fit rings out of 9),
%the ring model would not be too far off (7 good-fit rings out of 9),
while the ring and axisymmetric models would remain truly unacceptable 
(5 good-fit rings out of 9). 
The conclusions reached with the more rigorous consistency condition
(Eq.~(\ref{Eq_consist_prob})) are intermediate between the two situations 
described above: with the imposed probability set to $P_0=0.1$ ($P_0=0.05$), 
3 (4) rings can be properly fit with a ring magnetic field,
5 (5) with an axisymmetric field and 5 (7) with a bisymmetric field.

The quantitative differences between both criteria are easily understood.
A comparison between Eqs.~(\ref{Eq_consist_par}) and
(\ref{Eq_consist_prob}) immediately shows that the rule-of-thumb 
consistency condition (Eq.~(\ref{Eq_consist_par})) corresponds to
a probability level $P_1$ such that $\chi^2_{\rm crit} (P_1) = n_i-\nu$,
or equivalently, $P_1 = P [\chi^2 > n_i-\nu]$.
With twice the original error bars on the observational line-of-sight
fields, the rule-of-thumb consistency condition would become 
$\chi^2 \le 4 (n_i-\nu)$, corresponding to a probability level $P_2$ 
such that $\chi^2_{\rm crit} (P_2) = 4 (n_i-\nu)$,
or equivalently, $P_2 = P [\chi^2 > 4 (n_i-\nu)]$.
The values of $P_1$ and $P_2$ as functions of $n_i-\nu$
are tabulated in Table~\ref{Table_P1}.
Clearly, the rule-of-thumb consistency condition with the original 
error bars implies high probability levels ($P_1 \sim 30\%-40\%$),
which make it overly difficult to satisfy; if the model under testing 
is correct, there is nonetheless a $\sim 30\%-40\%$ chance 
that $\chi^2$ exceeds $n_i-\nu$ and that the model will be rejected.
On the other hand, with twice the original error bars, the probability 
levels drop very low ($P_2 \sim 0.1\%-4\%$) and the consistency condition
becomes too easily satisfied; the risk is then to accept a model that 
is in fact incorrect.

\begin{table}
\caption{Probability levels for the rule-of-thumb consistency condition}
\label{Table_P1}
\centering
\begin{tabular}{c|c c}
\hline\hline
\noalign{\smallskip}
$n_i -\nu$ & $P_1 = P [\chi^2 > n_i-\nu]$ & 
$P_2 = P [\chi^2 > 4 (n_i-\nu)]$ \\
\noalign{\smallskip}
\hline
\noalign{\smallskip}
1 &  0.3173 & 0.0455 \\
2 &  0.3679 & 0.0183 \\
3 &  0.3916 & 0.0074 \\
4 &  0.4060 & 0.0030 \\
5 &  0.4159 & 0.0012 \\
\noalign{\smallskip}
\hline
\end{tabular}
\end{table}

The bottom line is that the standard rule of thumb is way too stringent,
while the rule of thumb with twice the original error bars is way too easy.
In contrast, the more rigorous criterion with a probability level set to 
$P_0=0.1$, intermediate between $P_1$ and $P_2$, provides a reasonable 
trade-off.
The latter criterion is also more trustworthy, insofar as all the rings 
for all the models are tested with the same probability level.

The results of the present study suggest that the true large-scale 
magnetic field in our Galaxy has a more complex configuration than 
a strictly axisymmetric or bisymmetric field.
One possibility is that it consists of the superposition of axisymmetric,
bisymmetric and probably higher-order azimuthal modes.
Such combinations of modes have been observed in a number of external
galaxies \citep[e.g.,][]{Beck96,Berkhuijsen97,Rohde99,Beck07}.

%----------  acknowledgements   ----------------------------
\begin{acknowledgements}
We thank the referee for his/her valuable comments as well as 
Pierre Jean and William Gillard for useful discussions.
JLH and MH were supported by the National Natural Science Foundation
(NNSF) of China (10521001 and 10773016) and the National Key Basic
Research Science Foundation of China (2007CB815403). MH was also
supported by the French Embassy in Beijing during her stay in Toulouse
in the framework of a bilateral co-supervised PhD program.
\end{acknowledgements}

%----------  thebibliography   ----------------------------

\end{document}